\documentclass[12pt]{article}
\usepackage{graphicx} % Required for inserting images

\usepackage{amsmath}
\usepackage{amssymb}
\usepackage{fullpage}
\usepackage{url}
\usepackage{amsthm}
\usepackage{graphicx}
\usepackage[section]{placeins}
\usepackage{comment}

\newcommand{\beq}{\begin{equation}}
\newcommand{\eeq}{\end{equation}}

\newcommand{\footremember}[2]{%
    \footnote{#2}
    \newcounter{#1}
    \setcounter{#1}{\value{footnote}}%
}
\newcommand{\footrecall}[1]{%
    \footnotemark[\value{#1}]%
}

\title{Role of Non-Exponential Reversal times in Aggregation Models of Bacterial Populations}

\author{%
Michael Batista\footremember{uh}{University of Houston, Department of Mathematics}%
\and Patrick Murphy\footremember{sjsu}{San Jose State University, Department of Mathematics and Statistics
}
\and Oleg A. Igoshin\footremember{rice}{Rice University, Departments of Bioengineering, of Biosciences, and of Chemistry;
Center for Theoretical Biological Physics and Rice Synthetic Biology Institute }%
\and Misha Perepelitsa\footrecall{uh}%
\and Ilya Timofeyev\footrecall{uh} \footnote{itimofey@central.uh.edu}%
}

\date{\today}

\begin{document}
\maketitle

\begin{abstract}
In this paper, we consider 1D agent-based and kinetic models of aggregation with reversals.
In particular, we fit a Gamma distribution to represent the run times in myxobacteria and 
analyze numerically the importance of non-exponential reversal times. We demonstrate that non-exponential reversal times aid aggregation and result in tighter aggregates. 
We compare and contrast the behavior of agent-based and kinetic models, and also consider kinetic models with aggregation driven by chemotaxis. Thus, incorporating non-exponential reversal times into models of aggregation can be particularly important for reproducing experimental data, such as aggregate persistence and dispersal.
\end{abstract}


%%%%%%%%%%%%%%%%%%%%%%%%%%%%%%%
\section{Introduction}
Living systems are known for their ability to organize spatially into complex structures. Such self-organization is a hallmark example of emergent behavior – the formation of complex patterns from simpler interacting components \cite{ref8,ref9,ref12,ref13}. Examples of spatial self-organization behavior include flocking of birds \cite{Bialek2012}, self-organization of insect colonies \cite{gordon_organization_1996,Buhl2006} as well as multicellular self-organization during tumor growth and wound healing \cite{deisboeck_pattern_2001,werner2003,werner2007}. Albeit somewhat less complex, the reorganization of bacterial biofilms such as those formed by \textit{ Myxococcus xanthus} \cite{Kaiser89,Kaiser03,Curtis07,Zhang10,Munoz16} is a premier model system to understand collective pattern formation. 

Mathematically, the modeling of pattern formation has been studied with both agent-based models \cite{vicsek95,Anderson2005,Sliusarenko2007Aggregation,Zhang2012mechanistic,Cotter2017Data-driven} and with formulation of the coarse partial differential equations (PDE) analogs \cite{toner98,tad2008,carrillo2010,Weber13,banasiak2013,degman2017,Perepel2022}. Starting from a microscopic agent-based description, the derivation of PDE models is often carried out in the limit of infinitely many agents using, for instance, kinetic theory. Such PDE models are then used in the analysis of the collective behavior, e.g. derivation of bifurcation parameters at which self-organization behaviors can occur.

In many agent-based and corresponding PDE models, it is assumed that the agents' behaviors are memoryless, e.g., the changes in their positions and other variables at the next time step only depend on these variables at a curent time and not on their prior values. In the same fashion, it is assumed that transitions between different agent states can be recast as continuous-time Markov chains. This, in turn, implies that the transition times in the system have exponential distributions. However, it has been recognized that many biological systems are not memoryless (e.g. \cite{grossman2016,nava2020,Zhang2018Agent-Based}), and as a result, times between various events are not exponentially distributed. Therefore, different models keeping track of the agents' internal clock also have been developed
(e.g. \cite{Erban2004,Erban2005,grossman2016,nava2020,Zhang2018Agent-Based}). 

For the aggregation and formation of rippling traveling waves in {\it M.  xanthus} colonies \cite{Kaiser03,Igoshin2001Pattern,Igoshin2004waves}, it has been recognized that these behaviors are controlled by spatial and temporal coordination of cell reversals when cells switch their head and tail and move in the opposite direction. It is critical to note that these reversals are of a periodic nature and that reversal times are not exponentially distributed. However, the effects of a non-exponential distribution for the reversal times on the aggregation patterns have not been systematically studied.
Mathematical agent-based models and corresponding PDE models for both exponential 
(e.g. \cite{Alt80, Othmer1988, hillenothmer1, hillenothmer2, Chalub2004, Erban2004, Erban2005, Degond2017, Degond2020})
and non-exponential reversals have been previously developed in the literature
(see also review \cite{eftimie2012} and references therein), and here we focus on the detailed role of the non-exponential reversals in the context of aggregation motivated by 
{\it M. xanthus}. In particular, we choose simulation parameters typical for their behavior and use an observational dataset to fit Gamma distribution for the distribution of reversals. We then compare and contrast the behavior of agent-based and kinetic models with exponential and non-exponential reversals.

In this paper 
we consider two types of models - an open-loop model and a closed-loop model. In the open-loop model, the location of the aggregate is given explicitly, and the position and the velocity of each agent determine whether this agent is moving towards or away from the aggregate. Then, reversal times are sampled from the corresponding distribution. For the closed-loop model, we consider the kinetic model coupled with an equation for chemoattractant. In this case, aggregation emerges due to the instability of the spatially-homogeneous profile (see e.g. \cite{gmnr2009}).  
We demonstrate numerically that non-exponential distribution of reversals accelerates aggregation and results in tighter aggregates in both models. The open-loop model is a simpler case-study that separates aggregation from the instability of a flat profile for the chemoattractant in the closed-loop model. Therefore, the underlying mechanism for producing tighter aggregates is the same in both types of models and is related to a smaller variable of the non-exponential distribution for reversals.

%%%%%%%%%%%%%%%%%%%%%%%%%%%%%%%
\section{Distribution of run times}
\label{sec2}
Several cell behaviors controlling cell aggregation in {\it M. xanthus} have been quantified and demonstrated to play a key role in agent-based modeling ( e.g. \cite{Cotter2017Data-driven,Igoshin2004waves,murphy_cell_2023,Zhang2018Agent-Based}). The behaviors include (i) cell alignment, (ii) changing to a stopped/slow-moving state at high cell density, and (iii) a bias in reversal times for cells going toward vs away from the aggregates. In the early aggregation stage, the reversal times are longer for cells going toward the aggregates and, therefore, speed up cell accumulation. In the current work, we will focus exclusively on the bias and explore the role of the distribution of cell reversal times. To this end, 
we use typical behavior {\it M. xanthus} to compare and contrast 
simulations of aggregation models with exponential and non-exponential distribution for the reversal times. Therefore, the main reason for using the {\it M.  xanthus} dataset for estimating parameters for the distribution of reversals is to study aggregation models in a realistic parameter regimes.
In particular, 
we employ a dataset of {\it M.  xanthus} cell behaviors that includes information about reversal frequencies and cell positioning relative to aggregates \cite{welchdata}.
This dataset is related to the previous publication \cite{murphy_cell_2023} and was obtained by postprocessing of the data in \cite{welchdata2}. Once cell trajectories were obtained using the postprocessing pipeline from \cite{murphycode}, they were split into segments based on when directional reversals occurred, with each segment having an associated run time measuring the time between reversals. We then calculated the angle $\phi$ of each cell relative to the nearest aggregate for each segment, and categorized cells as moving towards the aggregate or away based on the sign of $\cos(\phi)$.

\begin{figure}[ht]
\centerline{
\includegraphics[scale=1.2]{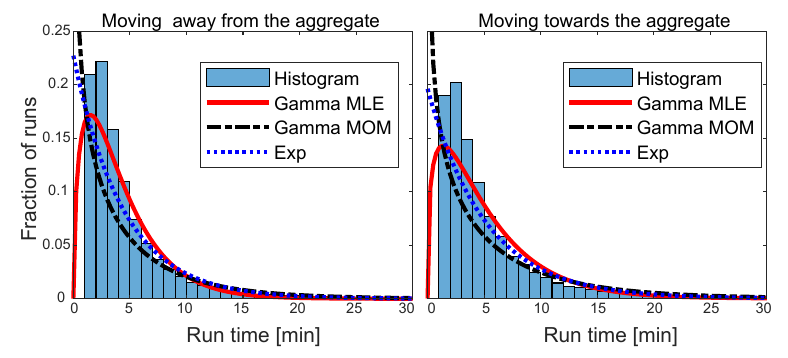}}
\caption{Histogram of run times for \textit{M. xanthus} moving towards and away from the aggregate, the Maximum Likelihood (MLE) fit of the Gamma distribution (used in this paper) for each case, the Method of Moments (MOM) fit for the Gamma distribution, and the Exponential fit.}
\label{fig1}
\end{figure}

We first estimate the statistical properties of the mean run times 
of \textit{M. xanthus}. In particular, we estimate the empirical means and variances of run-times away and towards the aggregate and fit exponential and gamma distributions to this data since it was demonstrated previously that gamma distribution results in a better fit for the reversal times \cite{grossman2016,Zhang2018Agent-Based}. 
%The means and variances are estimated 
%(separately towards and away from the aggregate) as
%\[
%\hat{m} = \frac{1}{N} \sum\limits_i \Delta \tau, \qquad
%\hat{v} = \frac{1}{N-1} \sum\limits_i \left(\Delta \tau - %\hat{m} \right)^2,
%\]
%where $N$ is the number of observations (towards or away) and $\Delta \tau$ is the run time (towards or away).
%
For the gamma distribution, we compared the Method of Moments and the Maximum Likelihood approaches.
Our tests (see Table \ref{tab2}) 
% TABLE WITH NORMS
%
\begin{table}[hbt!]
\centering
\begin{tabular}{|l|cc|cc|}
\hline
Test & MoM, Away & MLE, Away &  MoM, Towards & MLE, Towards \\
\hline
L1 & 0.57 & 0.24 &  0.56 & 0.26 \\
L2 & 0.25 & 0.11 &  0.31 & 0.11 \\
KL & 0.18 & 0.04 &  0.19 & 0.04 \\
\hline
\end{tabular}
\caption{Different error tests ($L^1$ and $L^2$ norms and the Kullback–Leibler divergence) comparing the accuracy of the Gamma distribution fits using the Method of Moments (MoM) and Maximum Likelihood (MLE) for the reversal times away and towards the aggregate.}
\label{tab2}
\end{table}
indicate that the Maximum Likelihood Estimators result in a better fit of the gamma distribution to the observational data. In particular, both the exponential distribution and the method of moments fit for the Gamma distribution severely overestimate the importance of very short reversal times.
Figure \ref{fig1} depicts experimental data and all three fits.

First, 
the empirical means of reversal times are estimated directly from data as averages of the run-times towards and away from the aggregate. The empirical means are approximately given by 
\begin{equation}
\label{mean}
\hat{m}_{toward} = 5.1, \quad
\hat{m}_{away} = 4.4.
\end{equation}
Next, we estimate the parameters of the Gamma distribution.
Using the Maximum Likelihood, the shape and scale parameters are estimated to be 
\begin{equation}
\label{partheta}
\begin{split}
& \theta_{toward} = 3.8, \quad
\theta_{away} = 2.7, \\
& k_{toward} = 1.36, \quad
k_{away} = 1.56.
\end{split}
\end{equation}
The corresponding first moments of the Gamma distribution are given by
\[
mean_{toward} = \theta_{toward} k_{toward} = 5.17, \quad
%mean_{in} = \theta_{in} k_{in} = 5.512, \quad
mean_{away} = \theta_{away} k_{away} = 4.21.
\]
These values are in close agreement with the mean run times \eqref{mean} computed from the dataset.
For the exponentially distributed reversal times we choose
\[
\theta_{toward} = 5.17, \quad
%\theta_{in} = 5.5241, \quad
\theta_{away} = 4.21
\]
which is consistent with mean values for the Gamma distribution of reversal times.
%

%%%%%%%%%%%%%%%%%%%%%%%%%%%%%%%
\section{Open Loop Model}
In the open loop model, agents are driven towards the center of the domain by prescribing different mean run times
(times between reversals) depending 
on whether agents are moving towards or away from the aggregate. In particular, the 
mean run times are given by
\[
mean_{toward} = 5.17 \, min, \quad
mean_{away} = 4.21 \, min.
\]

%%%%%%%%%%%%%%%%%%%%%%%%%%%%%%%
\subsection{Memoryless case}
\label{sec:openexp}
When the shape parameter $\kappa=1$,
the mean run times are exponentially distributed. This follows directly from the theory of continuous-time Markov chains 
(e.g. \cite{grimstir92,norris97}). This is also called the memoryless property because the length of the current run does not affect the probability of a reversal.
Thus, in this section we consider 
$k_{toward} = k_{away} = 1$
and $\lambda^{-1}_{toward} = mean_{toward}$, 
$\lambda^{-1}_{away} = mean_{away}$.

In addition, we also use typical values 
of the domain size (mm) and the velocity (mm/min)
(e.g. \cite{wolfgang2000, mauriello2010})
\[
x \in [-L, L] \text{~with~} L=1 \, mm, \quad 
v = 0.005 \, mm/min.
\]
Here we keep the velocity constant in magnitude, and
thus, one group of agents is moving to the right with the velocity $v_1 = 0.005$ and another is moving to the left with the velocity $v_2 = -0.005$. 
The motion of agents can be represented as a stochastic process
\begin{align*}
    x_i(t+\Delta t) &= x_i(t) + v_i \Delta t, \\
    v_i(t+\Delta t) &= W v_i(t),
\end{align*}
where $W$ is a random variable with 
$P(W=-1) = \lambda(x_i, v_i) \Delta t$, 
$P(W=1) = 1 - \lambda(x_i, v_i) \Delta t$, 
and $\lambda^{-1} =$ mean run time. Therefore, $\lambda(x_i, v_i)$ take only two possible values (towards and away) and can be defined as follows 
\[
\lambda(x_i, v_i) = 
\begin{cases}
    \lambda_{away} & \text{if~} v_i x_i > 0\\
    \lambda_{toward} & \text{if~} v_i x_i < 0.
\end{cases}
\]
For example, the right-moving particle (i.e. $v_i>0$), is moving towards the aggregate if $x_i<0$ and away from the aggregate 
if $x_i > 0$
and similarly for the left-moving group. 
We also employ periodic boundary conditions at $x= \pm L$.
%
% DEFINE LAMBDA BETTER

The corresponding kinetic model for the densities 
of two groups is 
\begin{equation}
\label{chem1}
\begin{split}
& \partial_t F+v \partial_x F = -h_+(x)F + h_-(x)G, \\
& \partial_t G-v \partial_x G =-h_-(x)G + h_+(x)F, 
\end{split}
\end{equation}
where $F(x,t)$ and $G(x,t)$ are densities of the right- and 
left-moving groups and $v > 0$. 
Equations \eqref{chem1}
are supplemented with periodic boundary conditions
$F(-L,t)=F(L,t)$ and
$G(-L,t)=G(L,t)$.
Functions $h_\pm(x)$ represent the hazard function of the exponential distribution and are defined as follows
\begin{equation}
\label{hpm}
h_+(x) = \lambda_+ = 1/\theta_+ , \qquad  
h_-(x) = \lambda_- = 1/\theta_- ,
\end{equation}
with
\[
\theta_+(x) = \begin{cases}
\theta_{toward} & x< 0\\
\theta_{away} & x>0
\end{cases}, 
\qquad
\theta_-(x) = \begin{cases}
\theta_{away} & x< 0\\
\theta_{toward} & x>0
\end{cases}.
\]

This model was first analyzed by Goldstein \cite{Goldstein51}
and then by others (e.g. \cite{othmer1998,hillen2000,Erban2004,Erban2005}). System \eqref{chem1} admits explicit solution for the steady state distribution 
$F(x,t) = G(x,t) = \rho(x)$
with
\beq
\label{rho}
\rho(x) = 
\begin{cases}
C  \exp(Sx) & x< 0  \\
C  \exp(-Sx) & x> 0
\end{cases}
\eeq
where $S=(b-a)/v$, $b=\theta_{away}^{-1}$, $a=\theta_{toward}^{-1}$, 
$C=\left(2/S (1 - \exp(-L S)\right)^{-1}$ is a normalization constant.

\begin{figure}[ht]
\centerline{\includegraphics[scale=1]{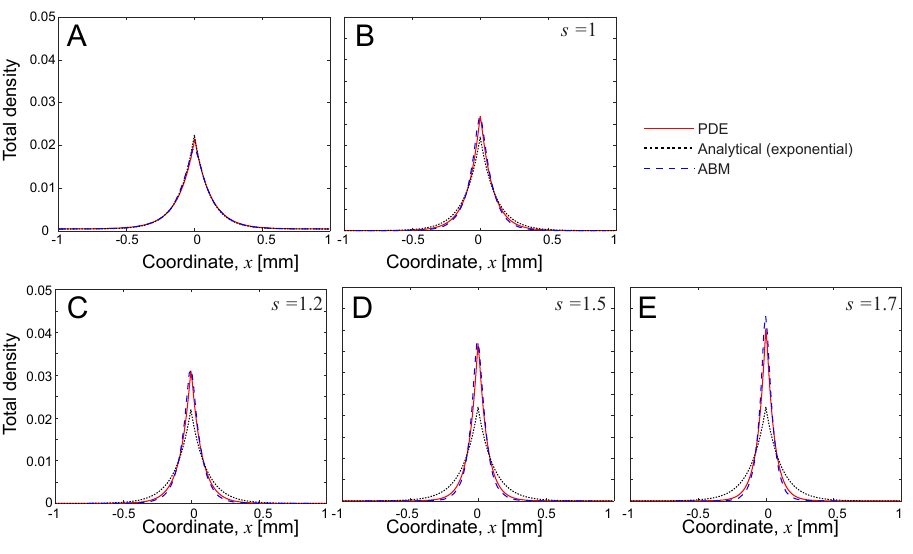}}
\caption{Simulations of the agent-based and kinetic open-loop models. Comparison of the total density at time $t=3000$ in simulations with the exponential (A) and non-exponential reversal times (B,C,D,E). 
The total density is defined as $F+G$ for the exponential case and $\int(f+h)\, d\tau$ for the non-exponential case.
Plot B - simulations with parameters \eqref{partheta} and $s=1$ in \eqref{eq:s}. C, D, E - simulations with parameters \eqref{partheta} and $s=1.2$, $1.5$, and $1.7$, respectively.}
\label{fig2}
\end{figure}

We perform numerical simulations to 
compare the long-time behavior of the agent-based model and the corresponding kinetic model. Initial conditions for the agent-based model are drawn from the uniform distribution and initial conditions for the PDE model also mimic the uniform distribution $F(x,0) = C_0(1 + 0.1\sin(2\pi x/L))$, $G(x,0) = C_0(1 + 0.1\cos(2\pi x/L))$, with the normalization $C_0 = 1/(4L)$ ensuring that the total density $F+G$ integrates to $1$. We also verified several other initial conditions, and the behavior described here is generic.The parameters of the agent-based model are: the total number of agents is $50,000$ and the time-step of simulation is $0.02$. 
In the simulations of the kinetic model, we rescale spatial units so that $L=200$ and $v=1$. Thus, the kinetic model is simulated using a limited Lax-Wendroff method with minmod limiter and $\Delta t=0.25$, $\Delta x=1$, $L=200$, $v=1$. The total time in both simulations is $T=3000$.
Comparison of the numerical solution and the analytical prediction $\rho(x)$ are depicted in 
Figure \ref{fig2}A where we plot the total agent density and the corresponding total kinetic density $F+G$.
As expected, we observe a very good 
agreement between the agent-based model and the kinetic model, and both solutions also agree very well with the analytical prediction.

%%%%%%%%%%%%%%%%%%%%%%%%%%%%%%%
\subsection{Non-Exponential Reversals}
\label{sec:nonexpopen}
Both
the agent-based model and the kinetic model discussed in section \ref{sec:openexp} can be extended to the case of non-exponentially distributed reversal times.
In this case, it is necessary to include the internal clock of each agent in the description of the stochastic dynamics. In particular, here we explicitly keep track of the run-time of each agent. Thus, the agent-based model becomes
\begin{align*}
    x_i(t+\Delta t) &= x_i(t) + v_i \Delta t, \\
    \tau(t+\Delta t) &=  
    \begin{cases}
    \tau(t) +\Delta t & \text{if } W=1 \\
    0 & \text{if } W=-1
    \end{cases}, \\
    v_i(t+\Delta t) &= W v_i(t),
\end{align*}
where
$P(W=-1) = h(x_i, v_i,\tau_i) \Delta t$, 
$P(W=1) = 1 - h(x_i, v_i,\tau_i) \Delta t$,
where $h$ is the hazard function of the corresponding Gamma distribution.
Here, $x_i$ and $v_i$ are only used to determine
whether the agent is right- or left-moving and select the appropriate parameters of the Gamma distribution. Denote the hazard functions for right- and left-moving agents as $h_+$ and $h_-$, respectively. Then, hazard functions are given by
\beq
\label{h}
h_+(\tau,x) {}={}\frac{p_+(\tau,x)}{1-\int_0^\tau p_+(\tau,x)\,d\tau}, \quad
h_-(\tau,x) {}={}\frac{p_-(\tau,x)}{1-\int_0^\tau p_-(\tau,x)\,d\tau} ,
\eeq
where 
\beq
\label{pplus}
p_+(\tau,x){}={}\frac{\tau^{k^+ -1}}{\Gamma(k^+)\theta_+^{k^+}}e^{-\frac{\tau}{\theta_+}},
\qquad
p_-(\tau,x){}={}\frac{\tau^{k^- -1}}{\Gamma(k^-)\theta_-^{k^-}}e^{-\frac{\tau}{\theta_-}},
\eeq
are the pdfs for the reversal times of right- and left-moving agents, respectively.
Parameters $\theta_\pm$ and $k_\pm$ are given by
\[
\theta_+(x) = \begin{cases}
\theta_{toward} & x< 0\\
\theta_{away} & x>0
\end{cases}, 
\qquad
\theta_-(x) = \begin{cases}
\theta_{away} & x< 0\\
\theta_{toward} & x>0
\end{cases},
\]
and
\[
k^+(x) = \begin{cases}
k_{toward} & x< 0\\
k_{away} & x>0
\end{cases}, 
\qquad
k^-(x) = \begin{cases}
k_{away} & x<0\\
k_{toward} & x>0
\end{cases}.
\]

The corresponding kinetic model is a system of PDEs for two populations $f(x,\tau,t)$ and $g(x,\tau,t)$
\beq
\label{chem2}
\begin{split}
\partial_t f + v\partial_x f+\partial_\tau f &= -h_+(\tau,x)f, \\
\partial_t g - v\partial_x g+\partial_\tau g &=-h_-(\tau,x)g,
\end{split}
\eeq
on the domain $x\in[-L,L]$ and $\tau\in[0,+\infty),$ with the boundary conditions
\[
f(-L,\tau,t){}={}f(L,\tau,t),\quad g(-L,\tau,t){}={}g(L,\tau,t), \quad \tau\geq0,\, t\geq 0,
\]
and 
\[
f(x,0,t) {}={}\int_0^\infty h_-(\tau,x)g(x,\tau,t)\,d\tau,\quad x\in[-L,L],\,t\geq0,
\]
\[
g(x,0,t) {}={}\int_0^\infty h_+(\tau,x)f(x,\tau,t)\,d\tau,\quad x\in[-L,L],\,t\geq0.
\]
Initial conditions for $f$ and $g$ also have to be provided. System \eqref{chem2} does not admit a closed-form stationary solution, thus we compare the agent-based and PDE models numerically. We also plot the analytical solution for the memoryless case.

Initial conditions for the agent-based model are drawn from the uniform distribution for the spatial variable, and exponential decay in $\tau$.
Initial conditions for the PDE model also mimic the uniform distribution in space and exponentially decaying in $\tau$. In particular, $f(x,\tau,0) = C_0 + 0.2C_0\cos(2\pi x/L) \lambda e^{-\lambda \tau}$, 
$g(x,\tau,0) = C_0 + 0.2C_0\sin(4\pi x/L + \pi/4)\lambda e^{-\lambda \tau}$, with $\lambda^{-1} = (mean_{toward} + mean_{away})/2$ and the normalization $C_0 = 1/(4L)$. The behavior described here is generic and emerges for other initial conditions as well.
Parameters in both agents-based and PDE simulations are similar to the ones used in the previous section, except the PDE model is integrated numerically using the Lax–Friedrichs method with $\Delta t=0.0125$, $L=200$, $\Delta x=0.5$, $\Delta \tau = 0.025$, $\tau_{max} = 80$. The fine discretization mesh in $\tau$ is necessary to accurately compute the integrals for the boundary conditions in $\tau$.

\begin{figure}[ht]
\centerline{\includegraphics[scale=1.0]{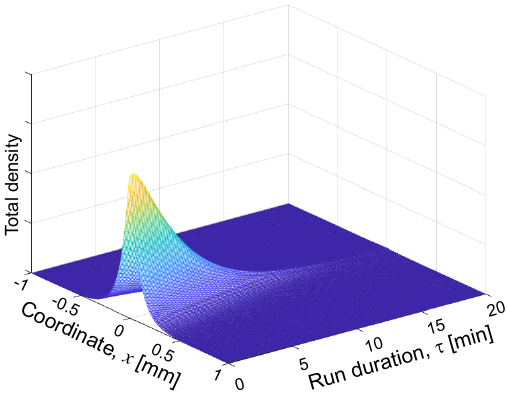}}
\caption{Simulations of the open-loop kinetic model \eqref{chem2} with non-exponentially distributed reversal times with parameters \eqref{partheta} and $s=1$. Density $(f+g)(x,\tau,t)$ at time $t=3000$ in the simulation of the kinetic model \eqref{chem2}.}
\label{fig4}
\end{figure}

  Figure \ref{fig2}B depicts the comparison of numerically computed densities for the agent-based model and the kinetic model with parameters \eqref{partheta}. For the kinetic model, we plot the total density 
$\int\limits_0^\infty (f+g) d\tau$.
Since there is no closed-form solution for the 
stationary density for the non-exponential case, we also plot the stationary density 
$\rho(x)$ in \eqref{rho} for the exponential case
with parameters $a=mean_{away}^{-1}$ and $b=mean_{toward}^{-1}$. 
The surface plot of $f+g$ versus $\tau$ and $x$ for $t=3000$
is presented in Figure \ref{fig4}.
We would like to point out that the means of all run-time distributions are identical in the exponential case and all simulations in this section. We also perform simulations with decreasing variance. In particular, we consider
\begin{equation}
\label{eq:s}
\theta_{\pm} = \theta_\pm / s, \qquad
k^{\pm} = s k^\pm .
\end{equation}
Recall, that the mean and variance of the Gamma distribution with parameters $(k,\theta)$ are
$k\theta$ and $k\theta^2$, respectively.
Therefore, the transformation above does not change the mean of the run times, but it decreases the variance for $s > 1$. The original parameters discussed in section \ref{sec2} are recovered with $s=1$.

Figures \ref{fig2}B-\ref{fig2}E depict the total density at the end of simulations ($t=3000$) for $s=1$, $1.2$, $1.5$, and $1.7$.

We observe a good agreement between the agent-based and PDE simulations (blue dash-dot and red solid lines in Figure \ref{fig2}) for all values of parameters. We also would like to comment that the PDE model reproduces the non-stationary behavior (not presented here for the brevity of the presentation) of the agent-based model very well. We also observe that the stationary density for the non-exponential case with $s=1$ (Figure \ref{fig2}B) is tighter than the stationary density for the exponential case (Figure \ref{fig2}A). Moreover,  
as the variance of the run-times is decreased (larger values of $s$; Figures \ref{fig2}C - \ref{fig2}E), the 
aggregate becomes even more confined.
%
%  Std Dev of the stationary profile
% exp = 32
% non-exp =27, 23, 19, 17
%
Corresponding standard deviations of the stationary profile are presented in Table \ref{tab1}. Therefore, we can conclude that a decrease in the variance of the distribution for run-times results in a decrease in the variance of the aggregate.
\begin{table}[hbt!]
\centering
\begin{tabular}{|l|ccccc|}
\hline
 & EXP & $s=1$ & $s=1.2$ & $s=1.5$ & $s=1.7$ \\
\hline
StdDev & $32$ & $27$ & 23 & $19$ & $17$  \\
\hline
\end{tabular}
\caption{Standard Deviation of the Stationary Profile for the agent distribution in simulations of the open-loop model for the Exponential Case (section \ref{sec:openexp}) and Non-Exponential Cases with $s=1$, $1.2$, $1.5$, $1.7$.}
\label{tab1}
\end{table}

\FloatBarrier

%%%%%%%%%%%%%%%%%%%%%%%%%%%%%%%
\section{Closed Loop Model}
In the closed-loop model, the reversals are driven by a chemotaxis model with the chemoattractant denoted as $u(x,t)$. Here we only consider PDE models of aggregation since we expect agent-based models to agree with the kinetic theory quite well (as demonstrated in the previous section).

%%%%%%%%%%%%%%%%%%%%%%%%%%%%%%%
\subsection{Exponential Reversals}
\label{sec:expclosed}
For the exponential reversals, the coupled model involves the kinetic model in \eqref{chem1} and the reaction-diffusion equation for the chemoattractant $u(x,t)$
\begin{equation}
\label{chem3}
\begin{split}
& \partial_t F + v \partial_x F = -h_+(x)F + h_-(x)F, \\
& \partial_t G - v \partial_x G =-h_-(x)G + h_+(x)F, \\
& \partial_t u = D \partial_{xx} u + \alpha (F + G) - \beta u,
\end{split}
\end{equation}
where parameters 
$D$, $\alpha$, and $\beta$ correspond to the diffusion, source, and decay of the chemoattractant. Equations \eqref{chem3} are supplemented with periodic boundary conditions for $F$, $G$, and $u$. 
A similar model was considered in \cite{gmnr2009}.
We consider 
$h_+(x) \equiv h_+(u_x)$
and 
$h_-(x) \equiv h_-(u_x)$
given by
\begin{align}
h_+(u_x) = & \lambda_+ = 1/\theta_+ \text{    with   }
\theta_+ = c \left(1 + \gamma_1 \tanh(\gamma_2 u_x) \right), \label{hpo} \\ 
h_-(u_x) = & \lambda_- = 1/\theta_- \text{   with   } 
\theta_- = c \left(1 - \gamma_1 \tanh(\gamma_2 u_x)\right), 
\label{hmo}
\end{align}
where $c, \gamma_1, \gamma_2 >0$. 
Here $h$ is the hazard function of the exponential distribution with mean $c = 1/\lambda$.
Parameter $c$ is the expected run time between reversals.
In this section we consider 
$c = (mean_{toward} + mean_{away})/2 \approx (5.2 + 4.2)/2 = 4.7$, 
$\gamma_1 = 0.1$, so that reversal times are in the range $[0.9c,1.1c] = [4.23,5.17]$, which corresponds to our simulations for the open-loop model.
 Thus, parameters are given by
 (here units are $mm$ and $min$) 
\[
L=1, \quad D=0.001, \quad \beta=0.1, \quad v=0.005, 
\]
\[
c=4.7, \quad \gamma_1 = 0.1, \quad \gamma_2 = 1.
\]
Therefore, the typical diffusion length scale is
$\sqrt{D/\beta} = 0.1$ and the averaged run length is $vc = 0.0235 \ll L$.
Parameter $\alpha$ plays the role of the bifurcation parameter. For smaller values of $\alpha$, we expect that the spatially independent profile of $u(x,t)$ should be stable, and for larger values of $\alpha$, we expect instability which leads to aggregation. 
Particular values of $\alpha$ are presented later in this section.
A similar PDE model was considered in 
\cite{gmnr2009} where
authors performed a bifurcation analysis with respect to the magnitude of the total density, $(F+G)$. Their analysis can be recast as bifurcations with respect to increasing 
parameter $\alpha$ while $\int (F+G) \,  dx$ is kept constant.
Computational parameters and initial conditions in the simulation of the closed-loop model 
\eqref{chem3} are $\Delta t=0.125$, $L=200$, $\Delta x = 0.5$,
\[
F(x,0) = C_0(1 + 0.1\sin(2\pi x/L)),
\quad
G(x,0) = C_0(1 + 0.1\sin(4\pi x/L + \pi/4)) 
\]
with $C_0 = 1/(4L)$ and
$u(x,0)=0.01$.

\begin{figure}[ht]
\centerline{\includegraphics[scale=0.8]{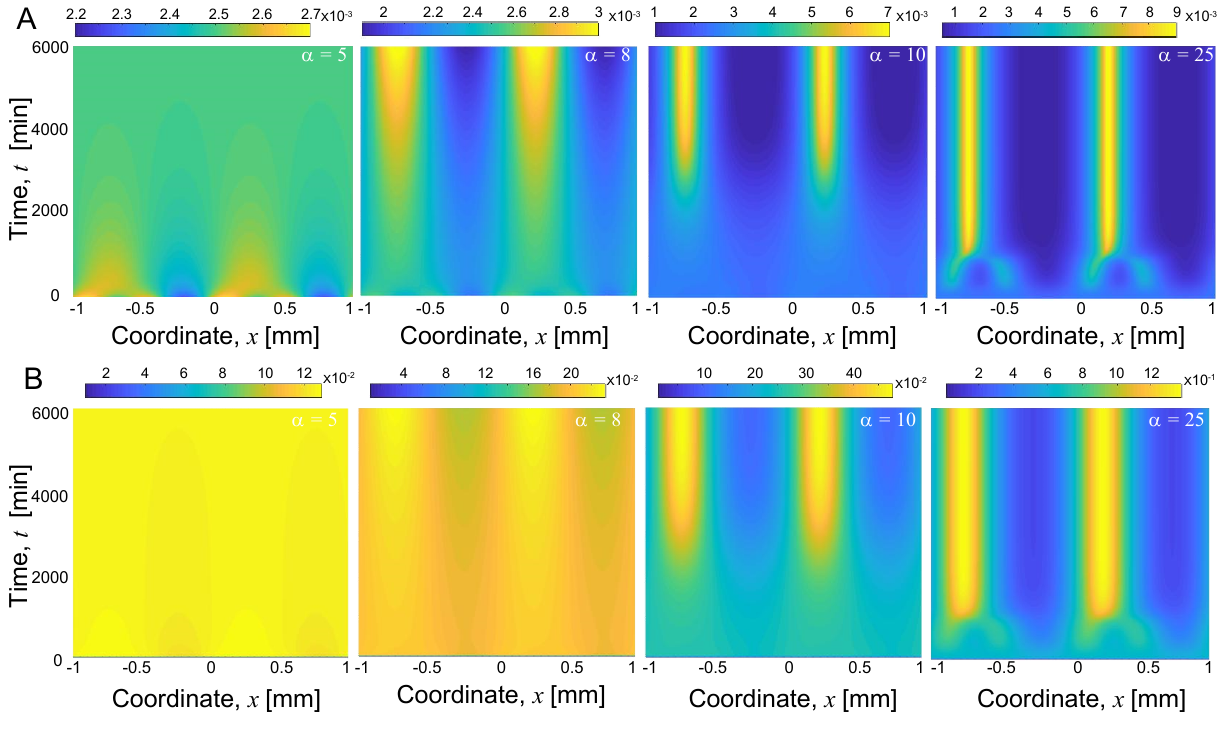}}
\caption{Simulations of the closed-loop model \eqref{chem3} with exponentially distributed reversal times for $\alpha=5$, $8$, $10$, $25$. Top row - total density
$(F+G)(x,t)$, bottom row - chemoattractant $u(x,t)$. Notice the difference in scales within the top row and the bottom row.}
\label{fig5}
\end{figure}

Total density $(F+G)$ and chemoattractant $u(x,t)$
for $\alpha=5$, $8$, $10$, $25$ are depicted in Figure
\ref{fig5} (parts A and B, respectively).
For $\alpha=5$, we observe that the spatially independent profile is stable (left column of \ref{fig5}A and \ref{fig5}B). Instability of the spatially-flat profile occurs at approximately $\alpha=7.5$, and for $\alpha=8$, $10$, $25$ we observe the formation of two aggregates. For $\alpha=8$, simulations do not seem to reach the stationary regime by the time $T=6000$, while for $\alpha=10$ and $\alpha=25$, the spatial profile appears to stabilize by $T=6000$. Also, for  $\alpha=25$, the profile stabilizes much faster compared to $\alpha=10$. Therefore, the strength of the instability clearly affects how fast the aggregation occurs.

The number of aggregates and their position depend on the initial conditions. We tested several different initial conditions, and we had simulations with a different number of emerging aggregates at short times. However, there is a characteristic length, and the system does not admit spatially stable profiles with too many aggregates (more than three in our case) over a longer time interval. When too many aggregates appear initially, they merge very quickly due to diffusion in the chemoattractant.
Moreover, even if there are only two aggregates, there seems to be a slow dynamics where the smaller aggregate is "absorbed" into a larger aggregate over a very long time. This occurs on a very slow diffusive time-scale and requires additional analysis and numerical investigation. It will be addressed in a consecutive paper.
%
%\begin{figure}[ht]
%\centerline{\includegraphics[scale=0.5]{fig_closed_exp/exp_alf5_den3.png}
%\includegraphics[scale=0.5]{fig_closed_exp/exp_alf8_den3.png}}
%\centerline{\includegraphics[scale=0.5]{fig_closed_exp/exp_alf10_den3.png}
%\includegraphics[scale=0.5]{fig_closed_exp/exp_alf25_den3.png}}
%\caption{Simulations of the closed-loop model \eqref{chem3} with exponentially distributed reversal times. Total density $(f+g)$ for $\alpha=5$, $8$, $10$, $25$.}
%\label{fig5}
%\end{figure}
%

\FloatBarrier

%%%%%%%%%%%%%%%%%%%%%%%%%%%%%%%
\subsection{Non-Exponential Reversals}
For the non-exponential reversal times, we couple the kinetic model \eqref{chem2} with the reaction-diffusion equation for the chemoattractant $u(x,t)$
\beq
\label{chem4}
\begin{split}
& \partial_t f+v_1\partial_x f+\partial_\tau f = -h_+(\tau,x)f, \\
& \partial_t g+v_2\partial_x g+\partial_\tau g =-h_-(\tau,x)g, \\
& \partial_t u = D \partial_{xx} u + \alpha (f + g) - \beta u.
\end{split}
\eeq
We would like to recall, that we consider the equation above 
on the domain $x\in[-L,L]$ and $\tau\in[0,+\infty),$ with the boundary conditions
\[
f(-L,\tau,t){}={}f(L,\tau,t),\quad g(-L,\tau,t){}={}g(L,\tau,t), \quad \tau\geq0,\, t\geq 0,
\]
\[
u(-L,t) = u(L,t), \quad t \ge0,
\]
and 
\[
f(x,0,t) {}={}\int_0^\infty h_-(\tau,x)g(x,\tau,t)\,d\tau,\quad x\in[-L,L],\,t\geq0,
\]
\[
g(x,0,t) {}={}\int_0^\infty h_+(\tau,x)f(x,\tau,t)\,d\tau,\quad x\in[-L,L],\,t\geq0.
\]
%

%\subsubsection{First Model}
The main goal of this paper is to investigate how fluctuations in the distribution of reversal times affect aggregation. For our parameter values the fluctuation in the means is $O(1)$ (see \eqref{mean}), the fluctuations in the scale parameter $\theta$ are also $O(1)$ (see \eqref{partheta}), and fluctuations in the shape parameter $k$ are $O(10^{-1})$ (see \eqref{partheta}). Also, the shape of the PDF of the gamma distribution depends weakly on $k$ in this range (see Figure \ref{fig1}). Therefore, we fix 
\[
k^* = 1.46
\]
and the model for $\theta$ becomes
\begin{align}
k^* \theta_+ = c \left(1 + \gamma_1 \tanh(\gamma_2 u_x) \right), \label{hpo2} \\ 
k^* \theta_- = c \left(1 - \gamma_1 \tanh(\gamma_2 u_x)\right).
\label{hmo2}
\end{align}
Thus, means of the Gamma distribution vary within the same range as in section \ref{sec:expclosed}.
Moreover, similar to section \ref{sec:nonexpopen} 
we also consider 
\begin{equation}
\label{closeds}
k^* = s k^*, \quad
\theta_\pm = \theta_\pm / s,
\end{equation}
so that we can study the effect of decreasing variance for $s>1$. 
%For $s=1.46^{-1}$, we recover the model for the exponential reversals considered in section \ref{sec:expclosed}.

Computational parameters are $\Delta t=0.25$, $L=200$, $\Delta x = 1$, $\Delta \tau = 0.05$, $\tau_{max} = 80$.
We use initial conditions that are identical in the spatial variable to those considered in the previous section and have an exponential decay in $\tau$, i.e.
\[
f(x,\tau,0) = C_0(1 + 0.1\sin(2\pi x/L)) \lambda e^{-\lambda \tau},
\quad
g(x,\tau,0) = C_0(1 + 0.1\sin(4\pi x/L + \pi/4)) \lambda e^{-\lambda \tau}
\]
with $C_0 = 1/(4L)$, $\lambda=1/c$, and
$u(x,0)=0.01$.
Other model parameters are identical to those in the previous section.

\begin{figure}[ht]
\centerline{\includegraphics[scale=0.8]{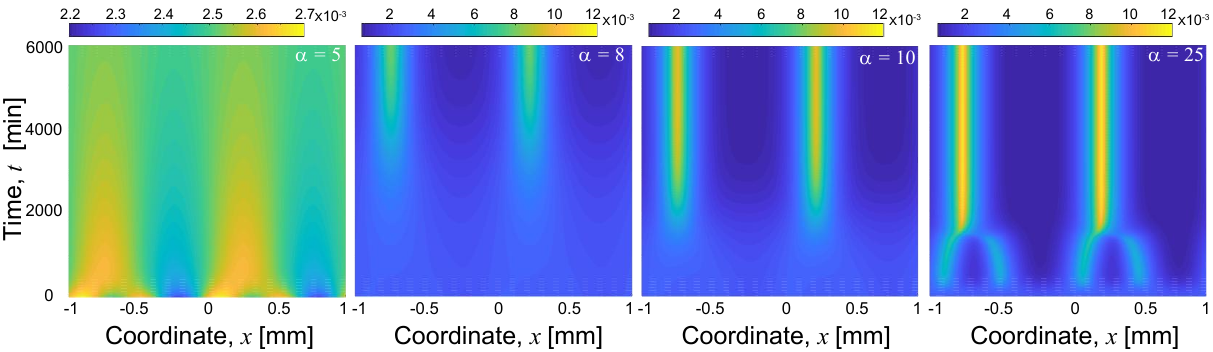}}
\caption{Simulations of the closed-loop kinetic model \eqref{chem3} with non-exponentially distributed reversal times with $k^*=1.46$. The color depicts the total cell density
$\int(f+g) d\tau$ for 
$\alpha=5$, $8$, $10$, $25$.}
\label{fig7}
\end{figure}

\begin{figure}[ht]
\centerline{\includegraphics[scale=1]{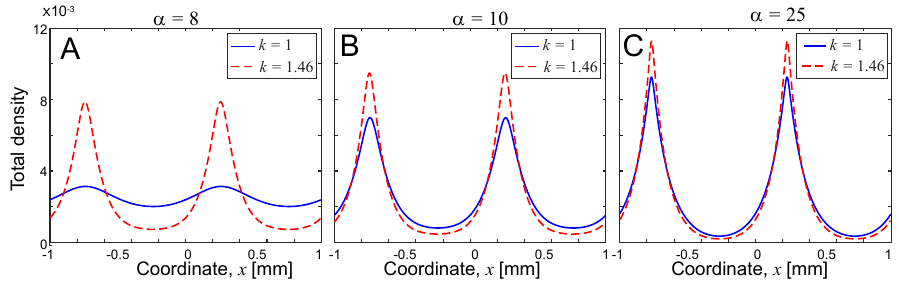}}
\caption{Comparison of the total density
at $T=6000$ in simulations of closed-loop kinetic models 
with the exponential ($k=1$; eq. \eqref{chem3}; Blue solid line) and 
non-exponential ($k=1.46$; eq. \eqref{chem4}; Red dashed line) distribution of reversals.
Parts A,B,C - $\alpha=8$, $10$, $25$, respectively.
The total density is $F+G$ and $\int(f+g) d\tau$ for kinetic models with exponential (eq. \eqref{chem3}) and non-exponential (eq. \eqref{chem4}) reversal times, respectively.}
\label{fig8}
\end{figure}

We discuss $s=1$ first.
Figure \ref{fig7} depicts the total density $\int (f+g) d\tau$
in simulations with non-exponential reversal times and 
$\alpha=5$, 8, 10, 25. 
These figures look qualitatively similar to the behavior of the total density with exponential reversal times (c.f. with the Figure \ref{fig5}A). In particular, the spatially flat profile is stable for $\alpha=5$. This profile becomes unstable for larger values of $\alpha$ and instability becomes stronger as $\alpha$ increases. However, there are important quantitative differences between simulations with exponential and non-exponential reversal times.
In particular, the non-exponential distribution of reversal times leads to tighter aggregates and faster aggregation. 
For both exponential and non-exponential reversals, simulations for $\alpha=10$ and $\alpha=25$ appear to reach the equilibrium state by the time $T=6000$ (Figure \ref{fig5}A and \ref{fig7}), but simulations with $\alpha=8$ still appear to be in the transient regime for both exponential and non-exponential reversal times.
We compare the total density profile at time $T=6000$ for simulations of closed-loop kinetic models with exponential and non-exponential reversals in Figure \ref{fig8}.
For $\alpha=10$, $25$ the difference in profiles is not very large. However, for $\alpha=8$ there is a considerable difference between the total density for the exponential and non-exponential reversals.
For the non-exponential reversals (Red line), the profile looks like two fully developed aggregates, but for the exponential reversals (Blue line), the profile looks like two very "weak" aggregates.
%
%\begin{comment}
%\end{comment}

\begin{figure}[ht]
\centerline{\includegraphics[scale=0.9]{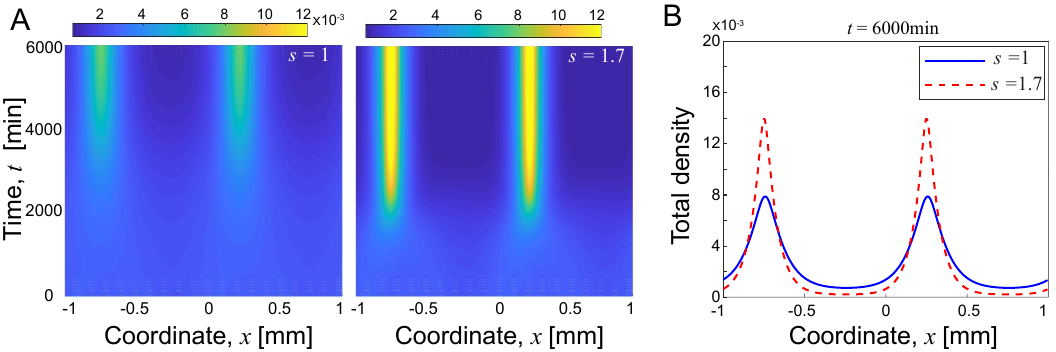}}
\caption{Comparison of the total density
$\int(f+g) d\tau$ in simulations of the closed-loop kinetic model \eqref{chem4} with non-exponential reversals with $k^*=1.46$ and $\alpha=8$.
Part A - comparison of the total density for $s=1$ (left) and $s=1.7$ (middle).
Part B - comparison of the total density at $T=6000$ for $s=1$ and $s=1.7$.}
\label{fig9}
\end{figure}

Next, we present results in simulations with a smaller variance $s=1.7$ in \eqref{closeds}. In particular, we consider the closed-loop non-exponential kinetic model with $\alpha=8$ and compare the total density for $s=1$ vs $s=1.7$ in \eqref{closeds} in Figure \ref{fig9}.
Since the mean of the Gamma distribution is $k\theta$,
transformation \eqref{closeds} does not affect the mean times of reversals.
However, the variance of the Gamma distribution is $k\theta^2$; thus, the variance is $1.7$ smaller for $s=1.7$ compared to $s=1$. 
Our results for the closed-loop kinetic model with non-exponential reversal times are consistent with 
previous results for the open-loop model (section \ref{sec:nonexpopen}) 
since a smaller variance for reversal times results in tighter aggregates. In addition, Figure \ref{fig9} also indicates that the aggregation occurs faster for $s=1.7$.
%
%
% \begin{figure}[ht]
% \centerline{\includegraphics[scale=0.5]{fig_closed/den_alf8s1_s1.7.png}}
% \caption{Comparison of the total density
% $\int(f+g) d\tau$ at time $t=6000$ in simulations of the closed-loop model with non-exponential reversals with $k^*=1.46$, $\alpha=8$ and $s=1$ (blue solid line) and $s=1.7$ (red dashed line).}
% \label{fig20}
% \end{figure}

%%%%%%%%%%%%%%%%%%%%%%%%%%%%%%%
\section{Conclusions}
In this paper, we consider several models of aggregation for agents with reversals and investigate numerically the importance of non-exponentially distributed run times (times between reversals). In particular, we consider open-loop models ABM and PDE models and closed-loop PDE models.
In open-loop models, all agents are "driven" towards the center of the aggregate by selecting the appropriate distribution (towards or away) for run times. In closed-loop models, dynamics for the agents' density is coupled to an equation for the chemoattractant and aggregation emerges due to the instability of a spatially flat profile. Notably, to ensure the models operate in biologically relevant regime, we employ an experimental dataset of {\it M. xanthus} 
in our study. Based on this data, the exponential fit for the run times severely overestimates the importance of short run times and a Gamma distribution provides a much better fit. It was previously shown that non-exponential reversal time distributions helped traveling wave patterns \cite{Igoshin2001Pattern, Igoshin2004waves} and we now have evidence it also aids aggregation.

The overall goal was to understand the role of non-exponential distribution for run times in the aggregation process of {\it M. xanthus}. Overall, our results indicate that models with exponentially and non-exponentially distributed run times produce qualitatively similar results. This can be expected for open-loop models since in this case all agents are "driven" towards the center of the aggregate. Comparison of closed-loop models is more elucidating since they are a system of complex equations for the agents' density coupled with an equation for the chemoattractant. Both closed-loop models (exponential and non-exponential) exhibit instability of the spatially flat profile that results in the formation of aggregates. 

However, there are some important quantitative differences between the behavior of the models with exponential versus non-exponential distributions of run times.
In particular, our numerical examples demonstrate that the aggregates for the non-exponential run-times are tighter and aggregation occurs faster compared with models with an exponential distribution of run times. This can potentially affect the size of aggregates and, thus, the distribution of aggregate sizes and their long-time dynamics. For instance, tighter aggregates are less likely to interact with each other, and thus, are likely to persist for longer times compared with wider aggregates in simulations with exponential distribution for run times. This suggests that non-exponential run times help prevent aggregates from dispersing prematurely. In {\it M. xanthus} aggregation, early aggregate are observed to disperse fairly frequently, and nearby aggregates can merge together, though not all do. \cite{Xie2011Statistical,murphy_cell_2023}. Non-exponential run times could be a mechanism to help prevent these phenomena, keeping the cellular aggregates persistent and distinct until the extracellular matrix that helps form the aggregate is fully developed.

%%%%%%%%%%%%%%%%%%%%%%%%%%%%%%%%%%%%
\section*{Acknowledgments}
I.T. and M.P. were partially supported by the grant 
NSF-DMS 1903270. OAI and PM were supported by NSF DMS-1903275 and IOS-1856742 (to O.A.I.).

%%%%%%%%%%%%%%%%%%%%%%%%%%%%%%%%%%%%%%%%
%\bibliographystyle{siam}
%\bibliography{refsall,refnew}

\end{document}